\documentstyle[11pt,aspconf]{article}
\def\aox {$\alpha_{\rm ox}$}
\def\ax {$\alpha_{\rm x}$}
\def\aouv {$\alpha_{\rm ouv}$}
\def\lopt {L$_{opt}$}
\def\lx {L$_x$}
\def\etal   {{\it et~al.}}
\def\lya {{Ly$\alpha$}}
\def\lyb {{Ly$\beta$}}

\begin{document}

\title{Emission Lines and the Spectral Energy Distributions of Quasars}
\author{B.~J.~Wilkes, P.~J.~Green, S.~Mathur \& J.~C.~McDowell}
\affil{Harvard-Smithsonian Center for Astrophysics, 60 Garden St.,
Cambridge, MA 02138, USA}

\begin{abstract}
Many years of study have failed to conclusively establish relations
between a quasar's spectral energy distribution (SED) and the emission lines
it is thought to produce. This is at least partially due to the lack of
well-observed SEDs.
We present initial results from a line--SED study for
a sample of 43 quasars and active galaxies for which we have optical and
ultra-violet spectra and far-infrared--X-ray SEDs. We
present the results of tests for correlations between line equivalent
widths and SED luminosity and slope parameters and compare these results
to those from earlier studies. We find that
the Baldwin effect is weaker when the luminosity is defined close to
the ionising continuum of that line and conclude that the detailed SED
is likely to be important in making further progress.
\end{abstract}


\section{Introduction}
It is generally believed that the strong, broad emission lines which
characterise quasar spectra are generated in gas photoionised by the
central continuum source. While this implies an intimate relationship
between the lines and the SED, many years of study have failed to
conclusively establish such relations. Indeed
there is a general dichotomy between
our qualitative viewpoints of the emission lines
and SEDs of quasars which belies a strong relation between the two. The
emission lines are thought to be similar from object to object,
and a single-zone model for the broad emission line region (BELR) was
used with success for many years (Netzer 1990).
In contrast the SEDs, including the ionising UV-soft X-ray continuum,
vary a good deal from one object to another (Elvis \etal\ 1994).
Possible explanations for this dichotomy include:
\begin{itemize}
\item The BELR gas sees the primary continuum, which has a fairly constant
shape, while our view is dominated by reprocessing in regions of the
quasar exterior to the BELR as well as along the line of sight.
\item A quasar's continuum is anisotropic so that the BELR sees a
different continuum from that seen by us.
\item Our viewpoints on the emission lines and SEDs of quasars are incomplete
and dominated by selection effects.
\item There exists a balance between the BELR gas and the continuum
such that the prominent emission lines have a small range
in relative strengths.
\item The emission lines are not emitted in photoionised gas.
\end{itemize}
Our sample of 43 quasars with complete SED and emission line information
allows us to address these possibilities. Here we present preliminary results
using standard single-band luminosities and slopes.

\section{The Sample}
Our sample consists of 43 quasars observed by the {\it Einstein}
X-ray satellite
with sufficient counts to define the X-ray spectral shape. It is a
heterogeneous sample, about half of which is radio-loud, and is biased
toward objects with relatively strong X-ray emission.
The multi-wavelength observations of the sample are described in
detail in Elvis \etal (1994).

For the purpose of this paper, ultra-violet (UV) spectra are taken
from the Lanzetta, Turnshek \& Sandoval (1993) compilation
of IUE spectra of quasars. The optical spectra were obtained by
ourselves on the MMT over the period 1985--1990 and will be presented
in Wilkes \etal\ (in preparation). Continuum luminosities \lopt, \lx\ and
\aox\ are taken from Wilkes \etal\ (1994), optical-UV slopes (\aouv) from Kuhn
(1996) and X-ray slopes (\ax) from Elvis \etal\ (1994) and Wilkes \&
Elvis (1987).
The UV and optical spectra were measured using automated routines
described in Green (1996). No deconvolution of heavily blended
features was attempted.

\section{Results and Discussion}

We investigated the relation between the equivalent (EW) width of
all the prominent UV and optical emission lines and the gross continuum
parameters: \lopt, \aox, \lx, \ax, \aouv.
In order to avoid biases introduced by weak or non-detected
lines, we use survival analysis techniques to include upper limits.
We applied the following tests for a significant correlation using the
ASURV package (LaValley, Isobe \& Feigelson 1992):
Cox Proportional Hazard Model,
Generalised Kendall's Tau, Spearman's Rho. The average probabilities of
a chance correlation derived from the results of these tests are presented in
Table 1. Values $<0.03$ are considered significant.

We find no correlation between the emission line EWs and
the slopes of either the X-ray or optical-UV continuum.
We find significant correlations between \aox\ and the UV lines
\lya+NV, OVI+\lyb~and CIV but not with CIII], MgII or the Balmer lines.
Correlations with the continuum luminosities are seen for all lines
except CIV. The traditional Baldwin effect (EW inversely related to \lopt,
Baldwin 1977) is the primary correlation for \lya+NV, MgII and the
Balmer lines while a similar correlation with \lx\ is primary for CIII]
and FeII$\lambda$4570. OVI+\lyb\ has too few data points (14, including
8 upper limits) to assess the correlations.
Similar results have been reported by
Green (1996), Zheng, Kriss \& Davidson (1995) and Corbin \& Boroson (1996)
respectively.

%
\begin{deluxetable}{lcccccc}
\tablewidth{0pt}
\tablecaption{Average Correlation Coefficients}
\tablehead{
\colhead{Line} &
\colhead{L$_{opt}$} &
\colhead{L$_{x}$} &
\colhead{\aox} &
\colhead{$\alpha_x$} &
\colhead{$\alpha_{ouv}$} &
\colhead{IC$^a$} \\
}
\startdata
OVI+Ly$\beta$ & NS$^b$ &0.06&0.03& NS & NS & X\\
\hline
Ly$\alpha$+NV & 0.006 & NS & 0.02 & NS & NS & O+X,X \\ 
\hline
CIV & NS & NS & 0.007 & NS & NS & O+X\\
\hline
CIII] & 0.04 & 0.015 & NS & NS & NS &O\\
\hline
MgII & 0.02 & NS & NS & NS & NS & X\\
\hline
H$\delta$ & 0.04 & NS & NS &-&-&O+X\\%
\hline
H$\gamma$+[OIII] & 0.04 & NS & NS &-&-&O+X\\
\hline
FeII4570 & NS & 0.01 & NS & NS & NS & X\\
\hline
H$\beta$ & NS & NS & NS & NS & NS &O+X\\ 
\hline
[OIII]5007 & 0.09 & NS & NS & NS & NS &\\ 
\hline
\enddata
\tablenotetext{a}{Ionising Continuum (Krolik \& Kallman 1988)}
\tablenotetext{b}{NS - correlation not significant, average probability of a
chance correlation $>$0.1}
\end{deluxetable}

The pattern of EW--luminosity correlations found here may
perhaps be understood if we refer to the part of the continuum believed
to be responsible for ionizing each emission line. The final column of
Table 1 indicates, very roughly, the spectral region to which a
line/blend is sensitive according to Krolik \& Kallman (1988). We define
the divide between optical (O) and X-ray (X) to be
0.01 eV, and find that, with the exception of the FeII
feature, the EW--luminosity inverse correlation is weaker with respect
to the luminosity defined closer to the ionising continuum for that line.
This confirms a similar result found by Green (1996) for a largely different
sample of quasars and implies that the ionising continuum is indeed
important in determining the strength of an emission line.
A more detailed investigation, including continuum luminosities
defined over narrower continuum bands, is likely to provide information
on the nature of this continuum in each case.
These results argue against a purely geometric explanation for the
Baldwin effect (eg. Netzer 1987) and in favour of scenarios involving
variations in the shape of the SED (eg. Mushotzky \& Ferland 1984).

FeII$\lambda$4570 is a notable exception to this general trend.
It is believed to originate in X-ray heated gas deep inside the
BELR clouds and so should be sensitive to hard X-rays, although
photoionisation models generally under-predict the line strengths, calling
their results into question.
A relation between the soft X-ray slope and the
FeII equivalent width, such that softer sources have stronger FeII,
in contrast to photoionisation model predictions, has
been debated for a number of years (Bergeron \& Kunth 1984,
Wilkes, Elvis \& McHardy 1987, Boroson 1989, Zheng \& O'Brien 1990).
Recent studies have suggested a complex set of relations which also include
line width and the strength of [OIII]$\lambda$5007
(Boroson \& Green 1992, Laor \etal\ 1994 \& 1996, Boller \etal\ 1996,
Lawrence \etal\ 1996). The presence of an inverse EW(FeII)--\lx\
correlation, which perhaps
suggests that the X-rays are not responsible for generating FeII,
along with the absence of a simple FeII--\ax\ correlation in the current sample
is intriguing and will be followed up in more detail.

\acknowledgments

BJW, PG, JCM gratefully acknowledge the financial support of
NASA contract NAS8-39073 (ASC) and SM of NASA grant NAGW-4490 (LTSA).
This paper is based in part on data obtained with the Multiple Mirror
Telscope (MMT), a joint facility of the Smithsonian Institution and the
University of Arizona.

\end{document}